\documentstyle[multicol,graphicx,prl,aps]{revtex}

\begin{document}

\author{H.Srikanth, B. A. Willemsen, T.Jacobs and S.Sridhar}
\address{Physics Department, Northeastern University, 360 Huntington Avenue, Boston,\\
MA 02115}
\author{A.Erb, E.Walker and R.Fl\"{u}kiger}
\address{DPMC, Universit\'{e} de Gen\'{e}ve, CH-1211 Gen\'{e}ve 4, Switzerland}
\title{New features in the microwave response of $YBa_2Cu_3O_{6.95}$ crystals :\ Evidence for a
multi-component order parameter }
\date{September 4, 1996}
\maketitle

\begin{abstract}
New features are reported in precision measurements of the complex microwave
conductivity of high quality $YBa_{2}Cu_{3}O_{6.95}$ crystals grown in $%
BaZrO_{3}$ crucibles. A {\em third} peak in the normal conductivity, $\sigma
_{1}(T)$, at around $80K$, and enhanced pair conductivity $\sigma_{2}(T)$ below $\sim 65K$ are observed. The data are inconsistent with a {\em %
single} order parameter, and instead are indicative of multi-component superconductivity. Overall, these results point to the presence of multiple pairing
interactions in $YBa_{2}Cu_{3}O_{6.95}$ and also provide a natural
explanation to account for the low temperature $35K$ conductivity peak
observed in all $YBa_{2}Cu_{3}O_{6.95}$ crystals.
\end{abstract}


\begin{multicols}{2}
The mechanism for superconductivity and the nature of the superconducting
state in the cuprates continue to be prime issues that are being debated 
\cite{JRKirtley95a}. Microwave measurements have yielded important
information on the nature of the pairing, the quasiparticle density of
states and scattering in the cuprate superconductors \cite
{WNHardy93a,TJacobs95d,SFLee96}. Surface impedance of $YBCO$ single crystals in the past have consistently shown two features: A linear penetration depth ($\lambda (T)\propto T$) over a limited
low $T$ range, and the presence of a bump in the surface resistance $R_s(T)$ which results in a peak in the normal conductivity $\sigma_1(T)$ at $\sim 35K$ well below the superconducting $T_c \sim 93K$ \cite{WNHardy93a,TJacobs95a}. The former behavior has been attributed to $d-wave$ order parameter symmetry or in general, the presence of nodes in the gap. A rapidly decreasing quasiparticle scattering rate below $T_c$ has been proposed to account for the latter feature \cite{TJacobs95a,DABonn92a}. There is a general consensus that the experimental results can be explained in the framework of a single $d-wave$ order parameter with inclusion of effects due to strong
coupling, scattering with strong temperature dependence, and
fluctuations\cite{TJacobs95a,PJHirschfeld94a}.

In this paper, we present new results on the microwave response of high
quality $YBa_2Cu_3O_{6.95}$ single crystals grown by a new method which
avoids crucible corrosion. We measure the temperature dependent surface impedance $Z_s=R_s+iX_s$
and penetration depth $\lambda $ ($=X_s/\mu _0\omega $) from which we extract the complex conductivity $\sigma
_s=\sigma _1-i\sigma _2$.  The two features mentioned above,  $\lambda(T)\propto T$ at low $T$ and the $35K$ peak in $\sigma_1$ are still present in these crystals. However in addition, a new peak in $\sigma_1(T)$ at $\sim 80 K$ and a distinct increase in $\sigma_2(T)$ below $\sim 65 K$ are observed.  A single $d-wave$ order parameter is insufficient to describe the new data, and instead we show that an analysis in terms of two-component superconductivity is necessary. 
  
The single crystals (typically $1.3\times 1.3\times 0.1mm^3$ in size) used
in this work were of very high quality grown from $BaZrO_3$ crucibles $(BZO)$%
\cite{AErb95a}. This growth method leads to crystals with extremely clean
surfaces and exceptional purity exceeding $99.995\%$\cite{AErb96a} in
contrast to crystals grown in other crucibles which have final reported
purities of $99.5-99.95\%$\cite{HIkuta96,HCasalta96}. As shown in this
paper, this difference in purity appears to play a significant role in the
microwave properties. Standard oxygen annealing procedures were followed to
obtain optimally doped crystals with oxygen stoichiometry around $O_{6.95}$ \cite{AErb96b}.
The crystals have $T_c=93.4K$, and very sharp transitions in $R_s(T)$ and
SQUID magnetic susceptibility measurements. Crystals grown by this method
also have unique physical properties due to their exceptional purity, as
revealed in other experiments \cite{IMaggioAprile95a,MRoulin96a}.
Unlike the case in earlier crystals, the vortex lattice was imaged for the first time in $YBCO$ with a low temperature STM \cite{IMaggioAprile95a}. Specific heat measurements indicated extremely sharp jumps at $T_c$ \cite{MRoulin96a}.

Three crystals of optimally doped $YBa_{2}Cu_{3}O_{6.95}$ grown in $BZO$
crucibles (hereafter called $YBCO/BZO$) were measured. In addition,
measurements on a $YBa_{2}Cu_{3}O_{6.95}$ crystal (of comparable dimensions)
grown in the commonly used yttria-stabilized zirconia($YSZ)$ crucible
(hereafter called $YBCO/YSZ$ ) are also presented for comparison.

The high precision microwave measurements were carried out in a $10GHz$ $Nb$
cavity using a ``hot finger'' technique \cite{SSridhar88b}. This method has
been extensively validated for precision measurements of the surface
impedance in cuprate \cite{TJacobs95d} and other superconductors. All
measurements reported in this paper were carried out with the microwave
field $H_{rf}\parallel \hat{c}$ so that currents flow predominantly in the $%
ab$ planes. Any influence due to $c-axis$ properties \cite{SFLee96} or finite size effects
are minimal as the results obtained on three samples with slightly different
dimensions and varying edge geometries were identical.

\begin{figure}[tbph]
\narrowtext
\begin{center}
  \includegraphics*[width=0.45\textwidth]{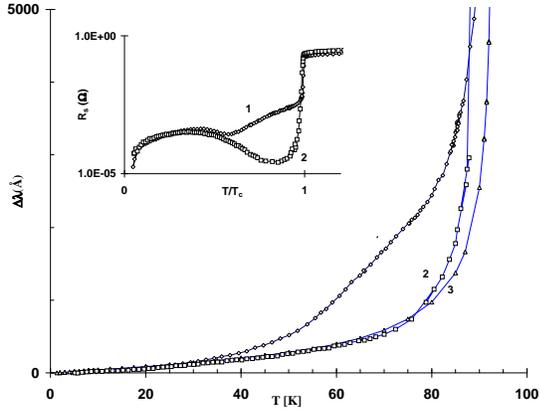}
\end{center} 
\caption{$\lambda (T)$ vs. $T$ for YBCO/BZO(curve 1) and YBCO/YSZ(curve 2) crystals. Data from ref.2 is also shown for comparison(curve 3). Lower inset shows the low T part. Top inset shows $R_s (T)$ vs. $T/T_c$ of YBCO/BZO and 
YBCO/YSZ.}  
  \label{fig1}
\end{figure}

The $T$ dependence of $\lambda (T)$ of $YBCO/BZO$ and $YBCO/YSZ$ crystals
are shown in Fig.\ref{fig1}. The data for the new high purity $YBCO/BZO$
crystals clearly reveal {\em a new feature} - a bump or plateau in the
vicinity of $60-80K$, which has not been reported in any previous $\lambda
(T)$ data for $YBa_{2}Cu_{3}O_{7-\delta }$ crystals. A plateau in $\lambda
(T)$ similar to that observed here has been reported in films and was
attributed there to a 2-gap behavior \cite{NKlein93a}.

The surface resistance $R_{s}(T)$ is also displayed in Fig.\ref{fig1} as a
function of $T/T_c$. At low temperatures all samples showed the non-monotonic
behavior in $R_{s}$ that is well-known and the low temperature values are
comparable to that seen previously \cite{TJacobs95a,DABonn92a}. However at
intermediate temperatures, a new feature is seen - a shoulder in $R_{s}$ in
the $YBCO/BZO$ sample which is completely absent in the $YBCO/YSZ$ sample.
We emphasize that the {\em data for }$YBCO/YSZ${\em \ is representative of
that reported in literature}\cite{TJacobs95a,WNHardy93a,SMAnlage96a}.

\begin{figure}[tbph]
\narrowtext
\begin{center}
  \includegraphics*[width=0.45\textwidth]{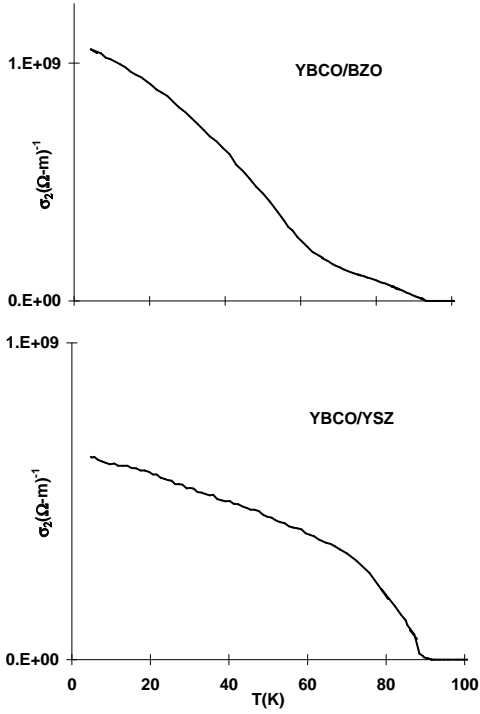}
\end{center} 
\caption{Pair conductivity $\sigma_2 (T)$ of YBCO/BZO (top) and YBCO/YSZ (bottom).}  
  \label{fig2}
\end{figure}

These new features are best studied in terms of the complex conductivity $%
\sigma _s=\sigma _1-i\sigma _2=i\mu _0\omega /(R_s+iX_s)^2$. The pair
conductivity $\sigma _2$ vs. $T$ is shown in Fig.\ref{fig2}, and is a measure of the superfluid density $n_s(T)$, since $\sigma
_2=n_s(T)e^2/m\omega =1/[\mu _0\omega \lambda ^2(T)]$ . Although only
changes $\Delta \lambda (T)\equiv \lambda (T)-\lambda (4.2K)$ and
correspondingly $\Delta X_s=\mu _0\omega \Delta \lambda $ are measured in
the experiment, the absolute value of $X_s$ and hence $\lambda $ are
obtained by assuming the normal state $R_n=X_n$ at and slightly above $T_c$.
This procedure yields a consistent value of $\lambda (0)\approx 1000\AA $
for all the $YBCO/BZO$ crystals and a value of $\lambda (0)\approx 1400\AA $
for the $YBCO/YSZ$ crystal. The $\sigma _2(T)$ data indicates an additional
onset of pair conductivity below around $60-70K$, never observed in previous
crystals reported to date as is evident from the comparison with the plot
for the $YBCO/YSZ$ crystal. Also the pair conductivity $\sigma _2(T)$ rises
to a higher value in the $BZO$ grown crystals compared with the $YSZ$ grown
crystals. All the $YBCO/BZO$ crystals showed this feature in the data. While
the accuracy of $\Delta \lambda \sim 1\AA $, the uncertainty in $\lambda (0)$
is much larger, perhaps of order $100-300\AA $. Nevertheless, the results
are consistent with a lower $\lambda (0)$ and hence enhanced pair
conductivity in the $YBCO/BZO$ crystals compared with the $YBCO/YSZ$
crystals.

\begin{figure}[tbph]
\narrowtext
\begin{center}
  \includegraphics*[width=0.45\textwidth]{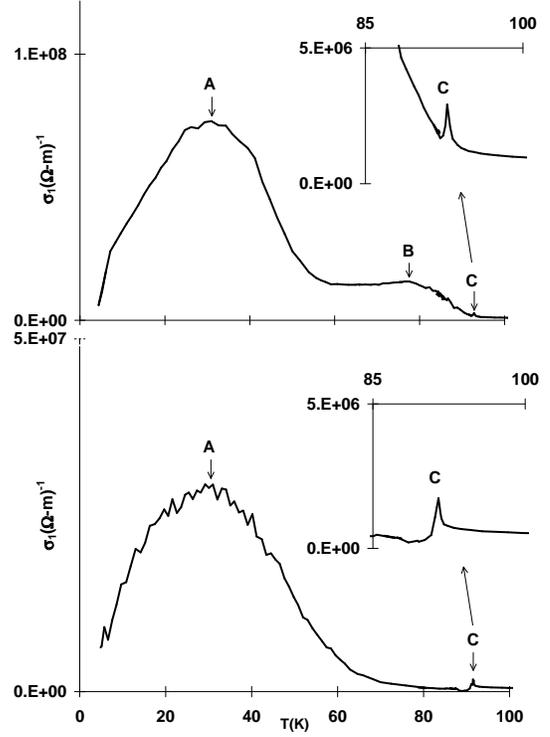} 
\end{center}
\caption{$\sigma_1 (T)$ of YBCO/BZO (top) and YBCO/YSZ (bottom). The sharp peaks C present in both cases are shown in insets. Note the appearance of a new peak B in YBCO/BZO.}  
  \label{fig3}
\end{figure}

The resultant plot of the normal conductivity $\sigma _1(T)$ is shown
in Fig.\ref{fig3}. The low temperature peak at around $35K$ (labeled $A$) is
observed in several previous measurements on crystals \cite
{TJacobs95a,DABonn92a}. Note that very near $T_c$ a sharp peak (labeled $C$)
is also present (see insets to Fig.\ref{fig3}) in all $YBCO/BZO$ and $%
YBCO/YSZ$ samples and which is often attributed to fluctuations \cite
{SMAnlage96a}. The sharpness of these peaks in the $YBCO/BZO$ crystals
attests to the high quality of these crystals. The most striking feature is
a new ({\em third}) conductivity peak (labeled $B$) clearly visible in the $%
YBCO/BZO$ data centered at approximately $80K$ ($\sim 0.9T_c$), which has
not been reported previously in microwave measurements.

It is clear that the $YBCO/BZO$ crystals reveal {\em two new and important}
features in the data - (i) the additional enhancement of pair conductivity $%
\sigma _2$ with an onset around $60-70K$, indicative of enhanced pairing
below this temperature, and (ii) the new {\em third} normal conductivity
peak at around $80K$ ($\sim 0.9T_c$) in $\sigma _1(T)$. The former was
observed in all the $YBCO/BZO$ crystals, while the $80K$ peak in $\sigma _1$ 
is more sensitive to sample details, particularly the normal state
scattering, and was lower in one of the samples which had the
highest $R_n$.(Details of results on the other $YBa_2Cu_3O_{6.95}$ and also
on specially oxygenated $YBa_2Cu_3O_{7.0}$ crystals will be published
elsewhere).

The data of Fig.\ref{fig1} for the new $YBCO/BZO$ samples {\em is difficult
to describe using a single order parameter below} $T_{c}$ . This is evident
from the curve for $\lambda ^{-2}(T)$ obtained from a weak-coupling d-wave
calculation shown in Fig.\ref{fig3}.

In the ``hydrodynamic'' limit, $\sigma _{1s,d}(T)=n_{qp}(T)e^{2}\tau (T)/m$.
A single peak in $\sigma _{1}(T)$ can arise from a combination of increasing 
$\tau $ and decreasing $n_{qp}$ as $T$ is lowered. Except for detailed
dependences on $T$, this peak should occur for an s- or d-wave
superconductor and even in a two-fluid model. (The BCS coherence peak for an
s-wave superconductor is masked by this effect). In order to describe the
conductivity data in earlier $YBa_{2}Cu_{3}O_{7-\delta }$ crystals \cite
{TJacobs95a,PJHirschfeld93a} in the framework of a d-wave model, one needs a
dramatic drop in the scattering rate below $T_{c}$. {\em However any model
using a single gap or order parameter (s- or d-wave) will only lead to a
single conductivity peak, in disagreement with the present data.}

\begin{figure}[tbph]
\narrowtext
\begin{center}
  \includegraphics*[width=0.45\textwidth]{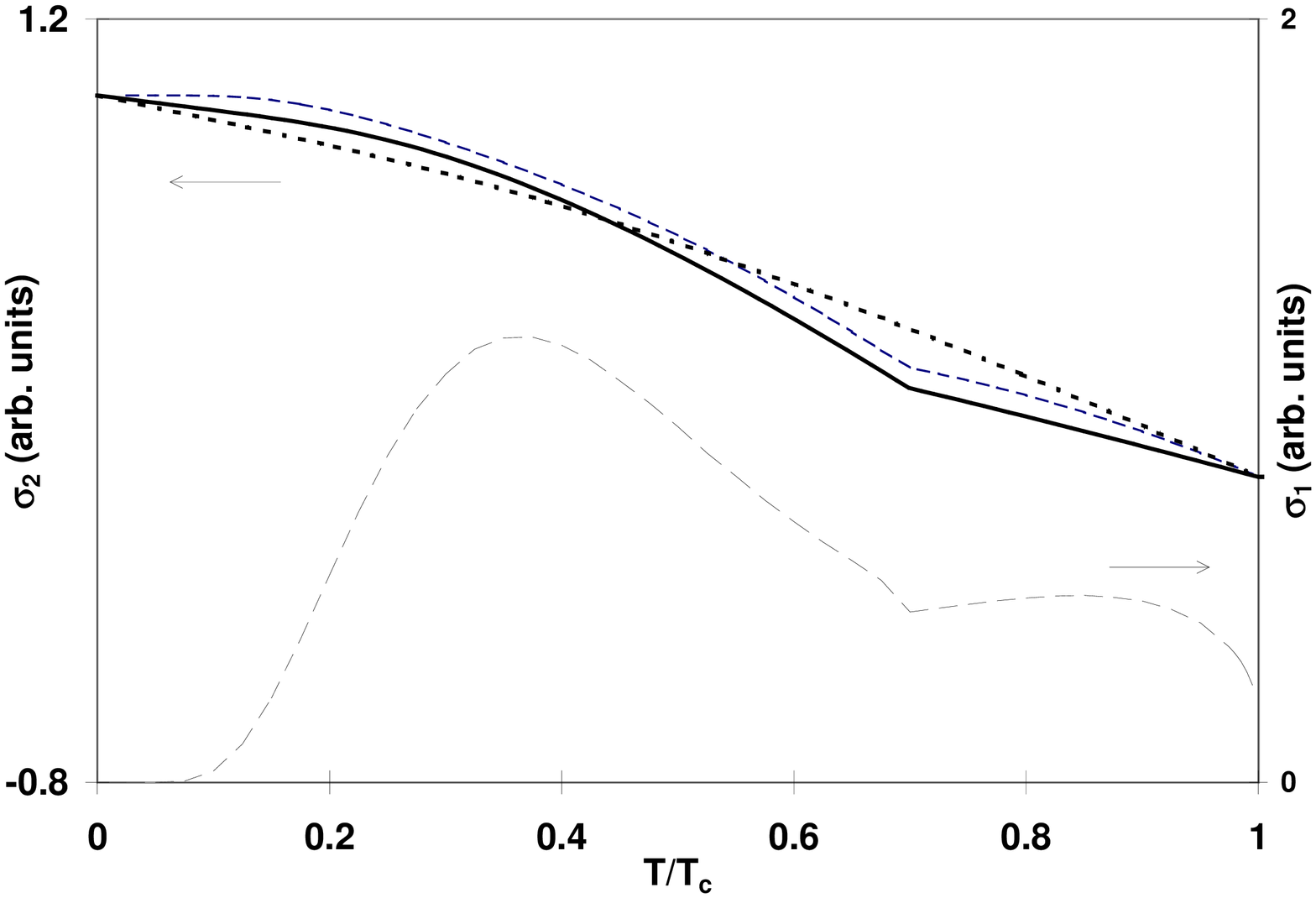}
\end{center} 
\caption{Calculations of $\sigma_1(T)$ and $\sigma_2(T)$ using the two component model assuming two $s-wave$ order parameters (long dashed lines). For $\sigma_2(T)$, the case for a more realistic situation of an $s+d$ symmetry is also shown (solid line). The short dashed line represents the calculations for a single weak-coupling $d-wave$ order parameter.}
  \label{fig4}
\end{figure}

Instead it is necessary to consider a two-component system to understand the
new data. The new feature in the $\sigma _2(T)$ data suggests that there is
additional pairing of carriers below approximately $60-70K$. The essential
features of the data can be well described by a simple model of two
superconducting components with $T_{cA}=65K$ and $T_{cB}=93K$. We have
calculated the total conductivity $\sigma _s\equiv \sigma _1-i\sigma
_2=(\sigma _{1A}+\sigma _{1B})-i(\sigma _{2A}+\sigma _{2B})$. For ease of
calculation we used s-wave order parameters for both the $A$ and $B$
components to calculate the conductivities using Mattis-Bardeen
expressions. In the case of $\sigma_2$, calculations assuming $s$ and $d$ which is closer to the realistic case are also presented. Gap values of $\Delta _A(0)/kT_{cA}=1.5$ and $\Delta
_B(0)/kT_{cB}=2.8$ were used. For $\sigma _1$, temperature dependent
scattering times $\tau _{A,B}=\tau _{oA,B}/(1+(T/T_{A,B}^{*})^4)$, with $%
\tau _{0A}/\tau _{0B}=2,T_A^{*}=37K,T_B^{*}=83K$ were assumed. The
calculations reproduce the essential features of the data , viz. the onset
of pairing around $60K$ observed in $\sigma _2(T),$and the 2 peaks in $%
\sigma _1(T)$, as shown in Fig.\ref{fig4}. Although we have used two
decoupled components in the above model for illustrative purposes, a small
attractive coupling between the two components is probably essential. It is
easy to show in the framework of Ginzburg-Landau theory \cite{JBetouras95a}
that such coupling does not change the results much, and the essential
features of the data are retained in more elaborate models.

The microwave measurements do not directly yield the symmetry of the order
parameter in the $A$ and $B$ components, except as can be inferred from the
temperature dependence of the data. The fact that $\lambda (T)$ is linear at
low temperatures would indicate that pairing in component $A$ could be
d-wave like, exhibiting nodes in the gap. Several theories have suggested
the strong possibility of a mixed $s+d$ state in the presence of
orthorhombic distortion as is the case in $YBCO$ \cite
{CODonovan95a,RCombescot95}, and detailed BCS calculations, such as those in
ref.\cite{CODonovan95a} of coupled s-d mixtures do yield penetration depth ($%
\lambda ^{-2}(T)$ vs. $T$) curves similar to the present $\sigma _2(T)$
data.

It is tempting to assign the two superconducting components $A$ and $B$ with
the associated condensates residing on $Cu-O$ planes and chains respectively
in the $YBCO$ system. However such an interpretation might be inappropriate
given the facts that the crystals are twinned and microwaves probe a length
scale spanning several hundred unit cells. Instead the two sets of
condensates may originate from different types of pairing associated with
different regions of the Fermi surface. Such a scenario based on band
structure taking into account the chain-plane coupling in $YBCO$ and
eventually leading to two types of pairing interactions has been proposed%
\cite{RCombescot95}.

Our results resolve an important issue regarding the origin of the
conductivity peak at $35K$. While in previous cases, it was necessary to invoke a precipitous drop in the scattering rate to account for its location, it is natural from Figs.\ref{fig2} and \ref{fig3} to{\em \ associate the
low }$T$ {\em (}$35K${\em ) peak with the A- (}$T_{cA} \approx 60K)$\ {\em component
and not the B (}$T_{cB} \approx 93K)${\em \ component}. The location at $35K$ ($\sim
0.5T_{cA}$) is now not unreasonable for a conductivity peak associated with
onset of pairing at $T_{cA}$. This implies that the type-$A$ condensate is
present in the $YBCO/YSZ$ samples also but has a weaker, almost gapless,
temperature dependence. We therefore believe that the same mechanisms were
operating in earlier samples also but are clearly distinguished in the new $%
YBCO/BZO$ crystals. 

A natural suspicion that arises is whether the data in the new $YBCO/BZO$
crystals arise from inadequate oxygen annealing leading to macroscopic
chemical phase separation. It is important to note that the annealing
procedures for the $YBCO/BZO$ crystals were exactly the same as for the $%
YBCO/YSZ$ crystals \cite{AErb96b}. An estimate suggests that the data cannot be explained
by macroscopic segregation of an $O_{6.5}$ $60K$ phase and an ideal $O_7$ $%
90K$ phase. For $O_{6.95}$ one would require $\sim 6-10\%$ of the $60K$
phase which cannot however account for the relative weights of the two
components in our $\sigma _2(T)$ data. Instead, our data points to a
physical mechanism like interlayer coupling rather than to chemical
segregation for our results.

In conclusion, measurements of the microwave properties of ultra-pure $%
YBa_{2}Cu_{3}O_{7-\delta }$ crystals reveal new features suggesting the
presence of two superconducting components in this compound. Our work
provides a possible explanation for the $35K$ microwave conductivity peak,
and yields new insights into the pairing mechanism in the high temperature
superconductors.

Work at Northeastern was supported by NSF-DMR-9623720, and at Geneva by the
Fonds National Suisse de la Recherche Scientifique. We thank R.S.
Markiewicz, A. Junod and J. Halbritter for useful discussions, D.P.Choudhury
and Z.Zhai for assistance.
\bibliographystyle{prsty}
\bibliography{big}

\end{multicols}
\end{document}